 \documentclass[twocolumn]{aastex6}

\usepackage{color}
\usepackage{multirow}

\newcommand{\bwt}{\begin{widetext}}
\newcommand{\ewt}{\end{widetext}}
\newcommand{\beq}{\begin{equation}}
\newcommand{\eeq}{\end{equation}}
\newcommand{\bea}{\begin{eqnarray}}
\newcommand{\eea}{\end{eqnarray}}

\begin{document}
                    
\title{Orbital Properties and  Gravitational Wave Signatures of Strangelet Crystal Planets}

\author{Jo\'as Zapata}
\author{Rodrigo Negreiros}
\affiliation{Instituto de F\'isica, Universidade Federal Fluminense, Av. Gal. Milton Tavares S/N, Niter\'oi, Brazil}


\begin{abstract}
In this paper  we consider the possibility that strange quark matter may be manifested in the form of  strangelet crystal planets. These planet-like objects are made up of nuggets of strange quark matter (SQM), organized in a crystalline structure. We consider the so--called strange matter hypothesis proposed  by Bodmer, Witten and Terazawa, in that, strange quark matter may be the absolutely stable state of matter. In this context,  we analyze planets made up entirely of strangelets  arranged in a crystal lattice. Furthermore we propose that a solar system with a host compact star may be orbited by strange crystal planets. Under this assumption we calculate the relevant quantities that could potentially be observable, such as the planetary tidal disruption radius, and the gravitational waves signals that may arise from potential star-planet merger events. Our results show that  strangelet crystal planets could potentially be used as an indicator for the the existence of SQM.  
\end{abstract}


\vspace*{5mm}
\section{Introduction}\label{sec:Int}
\label{sec:intro}


The hypothesis the strange quark matter (SQM), i.e. matter consisting of \textit{up}, \textit{down}, 
and \textit{strange} quarks is absolutely stable state of hadronic matter; has been proposed by 
Bodmer--Witten--Terazawa~\citep{bodmer1971collapsed, witten1984cosmic, terazawa1979ins}. 
If such hypothesis is true it would have implications on cosmology, the early universe, 
its development to the present day, astrophysical compact objects, and laboratory physics~\citep{crawford1993prediction, schaab1996thermal}. 
As of now, there is no sound scientific basis on which one can either confirm or reject it, 
so that it remains a possibility~\citep{schaab1996thermal, weber1996quark, glendenning1994strange}. 
Under such hypothesis one could consider that compact stars are, in fact, 
strange quark stars--self-bound objects composed of strange quark matter, 
as opposed to traditional, gravitationally bound neutron stars ~\citep{witten1984cosmic, haensel1986strange, alcock1986strange, glendenning1990fast}. Furthermore, in~\citep{witten1984cosmic, haensel1986strange,alcock1986strange, kettner1995structure, glendenning1995possible}, 
the authors further developed the concept of strange quark stars, that is, 
objects consisting of a strange quark matter core surrounded by a nuclear crust. 
They found a complete sequence of strange stars that range from very compact members, 
with properties similar to those of neutron stars, to white dwarf--like objects  (labeled strange dwarfs),  
to planetary--like strange matter objects or strange MACHOS. 
It was pointed out that the minimum--mass configuration in such sequence is $\sim 0.017M_{\odot}$, 
and that such values depend on the chosen value of inner crust 
density~\citep{kettner1995structure, glendenning1995possible}. 
If abundant enough in our Galaxy, such low--mass strange stars, whose masses and radii 
resemble those of ordinary planets could be seen by the gravitational microlensing 
searches~\citep{weber1996quark}. Furthermore, the authors of Ref.~\citep{alford2012strangelet} 
performed  a study on {\it strangelet dwarfs} -- which consist of a crystalline structure 
 of strangelets in a sea of electrons. In their work, they showed that if the surface tension 
 of the interface between strange matter and the vacuum is less than a critical value, 
 there is, at least, one stable branch in the mass-radius relation for strange stars. 

In the work presented here we follow the work of \cite{alford2012strangelet} - except we focus on 
planetary objects, characterized by low pressures.
Furthermore, recently there have been great advances in the detection of exo-planets
 - with a wealth of data available~\citep{perryman2000extra, borucki2016kepler, armstrong2016variability,schneider2011defining}. 
 Recently, Huang~\citep{huang2017searching} and Geng~\citep{geng2015coalescence}
  propose the use of such data to search for strange matter planets (in the context defined 
  by~\citep{kettner1995structure, glendenning1995possible}) by analysing two possible observational signatures:
   the tidal disruption radius and the gravitational waves (GW) emission from binary system composed by 
   a host compact star and a strange planet. They identified candidates of SQM planets by 
   using the following specifications: very small orbital period ($\leq 6100~s$), 
   an orbital radius smaller than $5.6\times 10^{10}~cm$ and,  
   possibly strong GW emission by strange planets with masses $\geq10^{-5}M_{\odot}$. 
   The idea is that, if any of these planets are, in fact, strange planets, 
   then due to their strange-matter properties (increased compactness, for instance) 
   their observational signatures should be different. 

In this work we follow in the footsteps of~\citep{huang2017searching, geng2015coalescence}, 
by searching for signatures of strange quark matter in the observed data of exo-planets. 
We consider, however, a somewhat more sophisticated model for strange quark planets, 
one which resembles more the actual structure of a planet. We recall that in the original proposal, 
strange planets were in fact very low mass strange stars (the order of a few Jupiter masses), 
with a small seed of strange matter in its core and relatively large nuclear crust extending all the way to the surface.
 In our model, as it will become clear, we consider the possibility of strangelets forming a crystalline structure, 
 such that, as in a white-dwarf, the objects are supported against gravitational collapse by electron degeneracy pressure.  Differently than in a white-dwarf (or the crust of neutron stars), in our proposed model there are no ions, 
 but rather, strangelets. Such objects could be formed by the same process that hypothetically 
 generated the compact star host (in a supernova, or stellar merger event, for instance) 
 with the high density matter giving birth to the high-density strange quark star (composed of homogeneous strange matter) 
 and lower density lumps, giving rise to crystalline strange planets. 


This paper is divided as follows: in section~\ref{Model}, 
we describe the microscopic model for  strangelet crystal planets, 
based on the ~\cite{Heiselberg1993} 
mass formula for the strangelet considering the screening of electric field,  from which, we can calculate the equation of state (EoS) of  strangelet crystal matter. 
With the EoS in hand, we compute the structure of the  strangelet crystal planets 
by imposing the appropriate hydrostatic equilibrium conditions. 
This sequence of planet--like objects is shown in section~\ref{SCP}. In section~\ref{Orbital} 
we look at the orbital properties of strangelet crystal planets and 
discuss possible observational implications. Finally, our discussions and conclusions are presented in Section~\ref{conc}.

\vspace*{5mm}
\section{Microscopic Model}\label{Model}
According to the SQM hypothesis, matter composed of roughly equal 
numbers of up (\textit{u}), down (\textit{d}) and strange (\textit{s}) quarks may 
be more stable than ordinary nuclear matter~\citep{bodmer1971collapsed, witten1984cosmic, terazawa1979ins}.
Farhi and Jaffe~\citep{farhi1984strange} have shown, within the framework of the MIT 
bag model, that for a wide range of parameters ($\alpha_{c}$ - 
the strong coupling constant, $m_{s}$, the mass of the strange quark and, $B$, the bag constant) 
strange matter could indeed be {\it absolutely} stable. If this is indeed the case, 
strange quark matter may be manifested in a wide range of ``sizes'', 
ranging from small {\it nuggets} with small baryon number $A$ -- 
analogous to nuclei -- all the way to bulk matter with very large $A$ - 
analogous to neutron stars. Based on this idea, Berger and
Jaffe~\citep{berger1987radioactivity, farhi1984strange} derived a mass formula  
for drops of strange matter -- in analogy to the Weizsaecker mass 
formula for nuclei. Later, this formalism was generalized 
by the authors of ref.~\citep{crawford1992production, desai1993conjectured}
- that have included a parameter to take into account the uncertainties in our understanding of surface effect.
 The mass formula for strangelets was further developed by the work of \cite{Heiselberg1993} who took into account the screening of electric fields. The efforts of \cite{Heiselberg1993} has also served as a foundation for the work of \cite{jaikumar2006strange} - who proposed, for the first time, that strangelets may be manifested in a crystalline structure. In here we follow the footsteps of the aforementioned authors - focusing on the low pressure regime, which is more appropriate for the planetary objects we seek to study.  
 For our study we have set $m_{s}=150$ MeV for strange quark mass and the bag constant is $B^{1/4}=146.96$ MeV.

As described in ref. \citep{Heiselberg1993} the energy per quarks in a strangelet may be written as 
\beq
E/N = \mu_0 + \frac{E_c + 4\pi \sigma R^2}{4\pi R^3 n/3}, \label{masstra} 
\eeq
where $\mu_0 = \mu_d = \mu_s = \mu_u$ (due to $\beta$ equilibrium), $R$ is the strangelet radius, $\sigma$ the surface tension, $n$ the quark density, and $E_c$ the Coulomb energy. In this work we are interested in crystalline strange matter, as such we need to restrict our study to the low surface tension regime. As discussed in \cite{alford2012strangelet} the critical value for the surface tension that allows the formation of a strangelet crystal is $\sim 1 - 10$ MeV/fm$^2$. In this work we use $\sigma = 0.6$ MeV/fm$^2$ - which is appropriate for the scenario we are interested. We note that we have also calculated strangelet properties for $\sigma = 0.2$ and $1.0$ MeV/fm$^2$ - which did not, significantly change the strangelet energy. By providing different values for $A$  we use eq. (19) of  \cite{Heiselberg1993} to calculate the charge of different strangelets, whose properties we present in Table \ref{table:strangelets}.  The strangelets we use in our study are appropriate to describe the low pressure/planetary objects we are interested in. One could can clearly see that, as expected, strangelets with higher $A$ 
have larger masses. Furthermore, the $Z/A$ ratio is small relative to  
ordinary nuclei as strangelets are less charged 
(due to the presence of the negatively charged strange quark)~\citep{madsen1995physical}.

%



\begin{table}[t]
\begin{center}
\begin{tabular}{|c|c|c|c|c|c|}
\hline
\multirow{2}{*}{Label} & \multirow{2}{*}{A} & \multirow{2}{*}{Z} & \multicolumn{3}{c|}{E ($10^{6}$ MeV)}         \\ \cline{4-6} 
                       &                    &                    & $\sigma =0.2$ & $\sigma =0.6$ & $\sigma =1.0$ \\ \hline
Stra$_{1}$             & $5\times 10^{3}$   & $581$              & $4.5189$      & $4.5189$      & $4.5201$      \\ 
Stra$_{2}$             & $1\times 10^{4}$   & $793$              & $9.0258$      & $9.0277$      & $9.0297$      \\ 
Stra$_{3}$             & $5\times 10^{4}$   & $1527$             & $45.055$      & $45.060$      & $45.066$      \\ 
Stra$_{4}$             & $1\times 10^{5}$   & $1986$             & $90.073$      & $90.082$      & $90.091$      \\ \hline
\end{tabular}
\end{center}
\caption{Properties of some strangelets from Eq.~(\ref{masstra}) for different constant surface tension values $\sigma$~(in $MeVfm^{-2}$).}
	\label{table:strangelets}
\end{table}
%
Once we have calculated the strangelet masses, we use a simplified Wigner-Seitz model, 
 much like in ref. \cite{alford2012strangelet}.
Thus, we consider a crystal structure of periodic spheres, in which, 
each cell of radii $R$ consists of a strangelet residing at its center, 
surrounded by an electronic cloud. Each cell will contains the amount 
of electrons needed to make it electrically neutral~\citep{glendenning2012compact,shapiro2008black}. 
Thus, for a given density $\rho$, the electron density is $\rho_{e} =(Z/A)\rho$, 
where $A$ is the mass number of the nuclear species and $Z$ is its atomic number. As we have mentioned before, we are interested in planetary objects - characterized by low pressure/density, thus the constant electron density approximation is appropriate \citep{alford2012strangelet}, we thus ignore the effect of pressure when calculating the strangelet properties, in particular Z(A). For stellar objects, where pressure effects cannot be ignored, one will have Z changing as a function of pressure/density due to electron capture.
We can then calculate the contribution of the strangelets, lattice and electron gas, 
obtaining the energy density and pressure given by
\beq
\epsilon(\rho)=\frac{\rho}{A} \bigg(E-Zm_{e}-\frac{9}{10}\frac{(Ze)^{2}}{R}\bigg)+\epsilon_{e}(k_{e}),\label{3}\\
\eeq
\beq
P(\rho)= P_{e}(k_{e})-\frac{3}{10}\bigg(\frac{4\pi}{3}\bigg)^{1/3}Z^{2/3}e^{2}\rho_{e}^{4/3}, \label{EoSCristalina}
\eeq
where $E$, is the mass of the strangelet, $e$, is the charge of electron and, 
$\epsilon_{e}(k_{e})$ and $P_{e}(k_{e})$, are the energy density 
and pressure of electrons, respectively~\citep{glendenning2012compact}. Note that for the low pressure regime we consider here, the energy density is mostly dominated by the strangelet contribution ($\rho E/A$).
The negative term on Eq.~(\ref{EoSCristalina}) is the lattice 
contribution to the total pressure. Furthermore, the third term 
on Eq.~(\ref{3}) represents electrostatic correction to the total energy density. 
Note that as in \cite{Heiselberg1993} we are using $e^2 = \alpha$,  the fine structure constant.
\begin{figure}[t]
	\centering
	\includegraphics[width=\columnwidth]{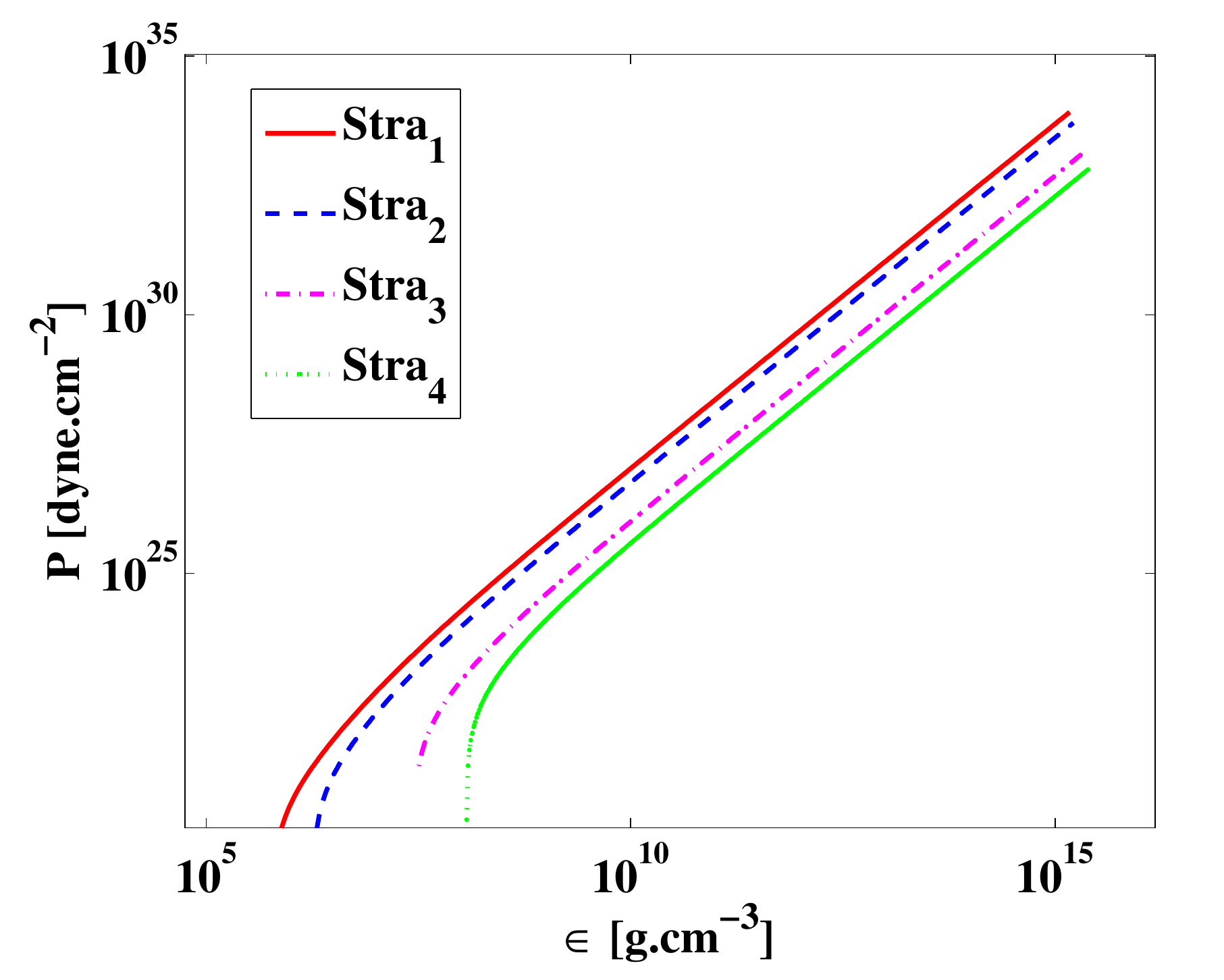}
	\caption{ Equations of state for   strangelet crystal planets made of different strangelet labeled as ``Stra'' from Table~\ref{table:strangelets} and Eqs.~(\ref{3})-(\ref{EoSCristalina}).}
	\label{fig:eos}
\end{figure}

The equation of state of such matter is shown if Fig.~\ref{fig:eos}, 
where we show the pressure as a function of energy density for the 
EoS associated with each of the studied strangeletes of Table~\ref{table:strangelets}.
Such results show that the crystalline matter proposed in this work 
has relatively low densities and pressure (in particular when compared 
to other works dealing with strange quark matter~\citep{alford2012strangelet}. It is worthwhile mentioning that  one gets a finite density at $P = 0$. As discussed in \cite{alford2012strangelet} this corresponds to the energy density of atomic cells, except in this case instead of ordinary atoms we have strangelet atoms, in which case the energy density at $P=0$  is significantly higher than that of ordinary atoms – which is explained by the fact that  “strangelet atoms” have a significantly higher mass ( $\sim 10^6$ MeV) than ordinary atoms, such as iron for instance, whose mass is $ 52 \times 10^3$ MeV.
Furthermore We note that less massive strangelets are associated with “stiffer” --EoS, i.e. higher pressures at the same energy density, which can be understood by the $Z/A$ ratio that is 
much higher for smaller strangelets, thus leading to denser electron gas - 
thus with a significantly stronger contribution to the pressure.

\vspace*{5mm}
\section{ Strangelet Crystal Planets} \label{SCP}
With the complete description of the microscopic physics 
(both of the strangelet and of the crystalling strange matter) 
we now proceed to determine the macroscopic properties of self gravitating objects. 
For this study we will consider spherically symmetric geometry, whose metric is given by 
\beq
ds^{2}=-e^{2\Phi}dt^{2}+ e^{2\Lambda}dr^{2}+r^{2}d\Omega^{2},
\eeq
where $\Phi$ and $\Lambda$ are functions of $r$ and, 
$d\Omega^{2}=d\theta^{2}+\sin^{2}\theta d\phi^{2}$. 
With the aid of Einstein's equation and assuming that the matter is a perfect fluid,
 one gets the  Tolman-Oppenheimer-Volkoff (TOV) equation, 
 that represents the hydrostatic equilibrium equation 
 for a spherical body~\citep{tolman1939static, oppenheimer1939massive}, given by
\beq
\frac{dp}{dr}=-\frac{(\epsilon(r)+p(r))(m(r)+4\pi r^{2}p(r))}{r(r-2m(r))},
\eeq
that needs to be solved in conjunction to the mass continuity equation,
\beq
\frac{dm(r)}{dr}=4\pi r^{2}\epsilon(r), \label{Massgravit}
\eeq
for a particular, EoS $P = P(\epsilon)$.
We now employ the EoS developed in section \ref{Model}, 
for a strangelet crystal configuration, to obtain the macroscopic 
properties of such objects. Since we are considering a solid, fully crystallize object, 
we found it appropriate to name them crystalline strange planets, so as to 
differentiate them from previous models of strange planets 
(in which strange quark matter is not crystallized - but present in bulk, 
in a small region of the object's core~\citep{kettner1995structure, glendenning1995possible, alcock1986strange, alcock1988exotic}). 

We are interested in the macroscopic properties of the planets, 
namely mass and radii; this is possible from the solution of  TOV's equation, 
and equation of state given by Eqs.~(\ref{3})-(\ref{EoSCristalina}). 
This system of differential equations is numerically integrated for 
a given central energy density $\epsilon_{c}$ from $r=0$ to $r=R$
 where the pressure vanishes, $p(R)=0$ (planet's surface). 
 Hence, we obtain the radius $R$ and the gravitational mass $M$ of the planet. 
\begin{figure}[t]
	\centering
	\includegraphics[width=\columnwidth]{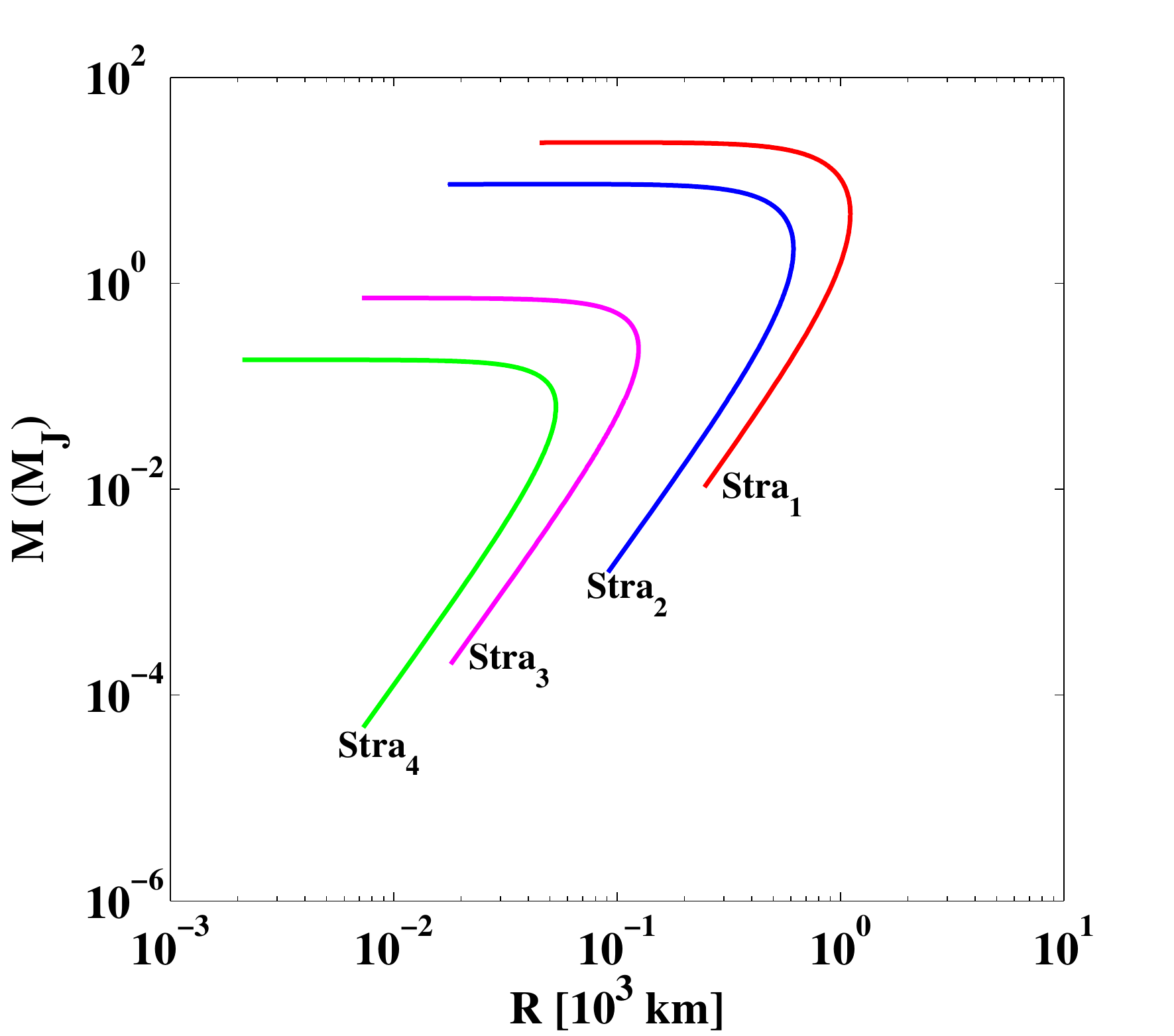}
	\caption{Mass versus radius of the  strangelet crystal planets for different strangelets from Fig.~\ref{fig:eos}.}
	\label{fig:family}
\end{figure}

The family of  Strangelet Crystal Planets is shown in Fig.~\ref{fig:family} 
where  the gravitational mass as a function of radius is exhibited. 
Given their low mass - and their planetary nature - the mass is given in Jupiter mass units. 
The results of Fig.~\ref{fig:family} show that, as expected, a stiffer EoS leads to planetary sequences with higher maximum mass. We can also note that for lower mass planets – associated with smaller central densities – one  sees that the mass is proportional to $R^3$ – which is typical for planetary objects which are mostly composed  of incompressible matter.
 An interesting result is that, contrary to what one would expect, crystals made of  heavier strangelets lead to a lower maximum mass for planets. Furthermore, for a given mass, planets made of lighter strangelets are larger, which is expected. 
The reason for such behavior has been already discussed -- heavier 
strangelets have a smaller $Z/A$ ratio, thus they have a 
less dense electron gas surrounding -- which yields to less pressure and hence lower masses.
\section{Orbital Properties of  Strangelet Crystal Planets}\label{Orbital}
Previous studies have tried to establish a possible connection 
between Strange Planets and exoplanets -- looking for possible 
aberrant behavior in observed exoplanets as possible evidence for strange quark matter.
Here we follow in the footsteps of ~\citep{huang2017searching} and~\citep{geng2015coalescence} - 
and determine the relevant properties associated with  strangelet crystal planets.

The first possibility we explore the orbit of exoplanets. 
Since strange planets are more compact, 
they can survive in closer orbits, where traditional hadronic planets
 would be subject to tidal disruptions. When a planet orbits around its host star, 
 the tidal force tends to tear the planet apart, but it can be resisted 
 by the self--gravity of the planet when the two objects are far from each other~\citep{gu2003effect}.
The critical distance, i.e. the so called tidal disruption radius ($r_{td}$), 
at which the tidal force is exactly balanced by the self--gravity of the planet, 
is defined as~\citep{hills1975possible}
\beq
r_{td}\approx\left(\frac{6M_{\star}}{\pi\overline{\epsilon}}\right)^{1/3}, \label{rtd}
\eeq
where $M_{\star}$ is the mass of the central host star and, $\overline{\epsilon}$ 
is the average density of the planet. If the distance is smaller than $r_{td}$, 
the tidal force will dominate and the planet will be completely broken up. 
Eq.~(\ref{rtd}) can be rewritten as 
\beq
r_{td}\approx1.5\times 10^{6}\left(\frac{M_{\star}}{1.4M_{\odot}}\right)^{1/3}\left(\frac{\overline{\epsilon}}{4\times 10^{14}}\right)^{-1/3}. \label{radii}
\eeq
One can also determine the period associated with orbits at the tidal 
disruption radius. From the Kepler's law, the radius and period 
of the orbit are related by~\citep{huang2017searching}
\beq
\frac{r^{3}}{P_{orb}^{2}}\approx\frac{GM_{\star}}{4\pi^{2}.}\label{orb}
\eeq

If we take $\overline{\epsilon}=30~g.cm^{-3}$ and $M_{\star}=1.4M_{\odot}$, 
the tidal disruption and the orbital period will be: $\sim5.6\times 10^{10}~cm$ 
and $\sim6100~s$, respectively. On the other hand, for  strange planets, 
with typical densities $\sim4\times 10^{14}~g.cm^{-3}$, we will have 
$\sim1.5\times 10^{6}~cm$ and ultra-short period $P_{orb}\sim 0.845~ms$. 
These results show us a possible way to identify SQM planets: 
if the tidal disruption and orbital period is significantly less than 
$\sim 5.6\times 10^{10}~cm$ and $\sim6100~s$, respectively, 
so it must be a strange planet~\citep{huang2017searching}. 
With these results, we can apply them to our model. 
\begin{table}[t]
	\begin{center}
		\begin{tabular}{|l|c|c|c|c|c|}
			\hline 
			Stra  & $M (M_{J})$ & $R (km)$ & $\overline{\epsilon}$ ($g/cm^{3}$) & $r_{td}$ (cm) &  $P_{orb}$ (ms) \\
			\hline
			1  &  $23.3$    &  $96$      & $2.5\times 10^{10}$ &  $3.8\times 10^{7}$  & 107 \\
			2  &  $9.2$    &  $53$      & $5.9\times 10^{10}$ &  $2.8\times 10^{7}$  & $69.9$\\ 
			3  &  $7.2$    &  $8.0$       &  $1.3\times 10^{13}$ & $4.7\times 10^{6}$ & $4.6$\\
			4  &  $0.181$  &  $3.65$ & $3.5\times 10^{12}$ & $7.3\times 10^{6}$ & $9.0$\\
			\hline
		\end{tabular}
	\end{center}
	\caption{Properties of maximum mass  strangelet crystal planets for each strangelet studied. Each strangelet is labeled by numbers. $M_{J}$, is Jupiter's mass ($M_{J}\sim 2\times 10^{-3}M_{\odot}$).}
	\label{table:proper}
\end{table}

\begin{table}[t]
	\begin{center}
		\begin{tabular}{|l|c|c|c|c|}
			\hline 
			$Stra$  & $M (M_{J})$ & $R (km)$ & $\overline{\epsilon}$ ($g/cm^{3}$) & $h$ \\
			\hline
			1  &  $23.3$     &  $96$      & $2.51\times 10^{10}$   &  $3.08\times 10^{-21}$  \\
			2  &  $9.2$     &  $53$      & $5.87\times 10^{10}$   &  $1.61\times 10^{-21}$  \\  
			3  &  $7.2$      &  $8.0$     &  $1.33\times 10^{10}$  & $7.68\times 10^{-21}$\\
			4  &  $0.181$  &  $3.65$     & $3.53\times 10^{12}$   & $1.24\times 10^{-22}$ \\
			\hline
		\end{tabular}
	\end{center}
	\caption{Gravitational wave amplitude for the crystalline strange planet merger events at the distance $d=10~kpc$. The host star has $M_{\star}=1.4M_{\odot}$.}
	\label{table:GW}
\end{table}

\begin{table*}[t]
	\begin{center}
		\begin{tabular}{lc c c c}
			\hline
			Object Planets & $\overline{\epsilon}$ & $r_{td}$ & $P_{orb}$ & $h$ \\
			& ($g/cm^{3}$)             & ($cm$)   & ($s$) &                 \\
			\hline
			Ordinary Planets \\
			\hline
			Low density  &  $10$    &  $5.1\times10^{10}$      & $\sim5263$ &  $4.9\times 10^{-29}$  \\
			High density  &  $30$    &  $5.6\times 10^{10}$      & $\sim6100$ &  $7.1\times 10^{-26}$  \\ 
			\hline
			 Strangelet Crystal \\
			Planets  &  $\sim10^{10}-10^{12}$    &  $\sim4\times 10^{6}-3\times10^{7}$       &  $\sim0.009-0.107$ & $\sim10^{-22}-10^{-21}$\\
			\hline
			Strange Planets  &  $\sim4.0\times10^{14}$  &  $\sim1.5\times10^{6}$ & $\sim8.45\times10^{-4}$ & $\sim10^{-23}-10^{-21}$ \\
			\hline 
		\end{tabular}
	\end{center}
	\caption{Comparisons:  strangelet crystal planets vs ordinary planets and strange planets.}
	\label{table:comparison}
\end{table*}

The orbital properties for the maximum mass  strange crystal planets found 
in our study  are shown in Table~\ref{table:proper}, where we have obtained 
the tidal disruption $r_{td}$ and $P_{orb}$ from Eqs.~(\ref{radii}) and~(\ref{orb}), 
respectively (using $M_{\star}=1.4M_{\odot}$). In that table, $M$, is the total mass of 
the planet, $R$, is the radius, $\overline{\epsilon}$, is the average density, $r_{td}$, is the tidal disruption  and, 
$P_{orb}$ is the orbital period. We note that the tidal disruption and the rotation 
periods of these planets are, as expected, much smaller than ordinary planets 
($5.6\times 10^{10}~cm$; $6100~s$). Also, we see that their densities are 
much higher when compared to ordinary planets ($\sim 1-30~g.cm^{-3}$).

Another aspect that warrants investigation is gravitational wave emission. 
According to General Relativity, orbital motion of a binary system can lead 
to Gravitational Waves emission and spiral--in of the system. 
Geng, Huang and Lu~\citep{geng2015coalescence}, showed that due to extreme 
compactness, strange planets can spiral very close to their host compact stars, 
without being tidally disrupted. These systems could potentially serve as a new source 
for Gravitational Waves (GW). GW emission from these events happening in our local 
Universe may potentially be strong enough to be detected by upcoming detectors such as 
Advanced LIGO~\citep{acernese2006virgo, abbott2009upper} and 
the Einstein Telescope~\citep{punturo2010third,hild2008pushing}.

This analysis can thus be used as possible evidence to  the existence of SQM. 
In contrast to normal matter planets moving around a compact star, their GW 
signals are negligibly small since the planet can not get very close to the central 
star due to the tidal disruption effect.

According to~\citep{geng2015coalescence}, the strain amplitude of GW 
from a strange--matter system, at the last stage of the inspiraling and 
at distance $d$ from us, is
\beq
h=l\left(\frac{M_{\star}}{1.4M_{\odot}}\right)^{2/3}\left(\frac{\overline{\epsilon}}{4\times 10^{14}}\right)^{4/3}\left(\frac{R}{10^{4}}\right)^{3}\left(\frac{d}{10}\right)^{-1}\label{hGW}
\eeq
where $l= 1.4\times 10^{-24}$, $d$ (in kpc) is the distance of the binary to us, and
$R$ is the radius of the star (in cm). If $r_{td}$ is too large, the GW emission will be very weak. 
For example, at $d=10~kpc$, for a typical planet with $M=5\times10^{-4}M_{Jup}$, 
$\overline{\epsilon}\sim 10~g.cm^{-3}$ and, $R\sim 3.6\times 10^{8}~cm$; 
disrupted at $5.1\times 10^{10}~cm$, the maximum GW amplitude is only $h\approx 4.9\times 10^{-29}$, 
screen for high density $\sim 30~g.cm^{-3}$, the GW amplitude is 
$h\approx 7.05\times10^{-26}$, which is too weak to be detected. 
For strange planets, however, the strain amplitudes of {\it GWs} are between 
$\sim10^{-23}-10^{-22}$, at a distance of $d\sim 10~kpc$~\citep{geng2015coalescence}. 

Now employing Eq.~(\ref{hGW}) to  strangelet crystal planets. 
We obtain the results shown in Table~\ref{table:GW}. 
From this table it is clear that  a binary system with crystalline strange planets 
can emit GW with amplitudes of the order of $\sim 10^{-22}-10^{-21}$ with frequencies ($=2/P_{orb}$) between $15-450$ Hz. This values are within of sensitivity of GW detectors like Advance LIGO and the Einstein Telescope \citep{geng2015coalescence}.

Finally, we can summarize the properties of strange crystal planets and 
compare them with the ordinary planets and strange planets. 
From the Table~\ref{table:comparison}, we see that our model differs from other
 cases in which: they have higher $r_{td}$ and $P_{orb}$ compared to strange planets, 
 and lower $r_{td}$ and $P_{orb}$ compared to ordinary planets. 
 Besides,  strangelet crystal planets and strange planets have the strain amplitude 
 in the same order, whereas, ordinary planets have a much lower 
 strain amplitudes than the other SQM objects.

\vspace*{5mm}
\section{Discussion \& Conclusions} \label{conc}
In this work we have proposed a novel planetary configuration --  strangelet crystal planets: low pressure objects made up of strangelets arranged in the periodic crystal~\citep{shapiro2008black, glendenning2012compact}.
This model differs from previously proposed strange star models ~\citep{kettner1995structure, glendenning1995possible, alcock1986strange, alcock1988exotic,glendenning1992nuclear,glendenning1990fast} -  
as in such models SQM is found in bulk and possibly surrounded 
by ordinary hadronic matter. In our study we follow the approach of \cite{jaikumar2006strange} in which we 
consider SQM in a crystalline structure. Within such scenario, self-bound strangelets can organize themselves 
in a crystal lattice, permeated by an electron gas (needed for charge neutrality); such matter could then form self-gravitating objects supported against gravitational pull by the electronic pressure, much like the white dwarfs do.

We have determined the masses and radii of such objects and found that their mass may be as high as that of ordinary planets  ($\sim 10^{-4}-25~M_{J}$) but with slightly smaller radii than ordinary planets. 
Furthermore, we have calculated possible observable signatures of such a model using 
the concept of tidal disruption radius and amplitude of gravitational waves that could be 
emitted by such systems. As expected we have found that, due to their compactness, 
the tidal disruption radii of strange crystalline planets is significantly smaller than that 
of ordinary planets. We have found however that when compared to previously proposed 
strange planet models - our scenario leads to higher tidal disruption radii. This means that 
 strangelet crystal planets exhibit an intermediate behavior, with possible orbital properties
not as extreme as that of strange planets, but not as mild as that of ordinary planets. 
This is not surprising, given the {\it ``hybrid''} nature of our proposal, which mixes properties 
of both traditional planets (solid, crystalline structure) with that of strange planets (strange quark matter).
As for the possibility of GW emission, we have found that our model does not 
differ significantly from the previously proposed strange planets - with the amplitude of 
GW from both models being somewhat similar. However both models predict a GW 
amplitude significantly higher than that of ordinary planets which could potentially be 
detected by the future Einstein Telescope or the Advance LIGO~\citep{geng2015coalescence} - 
which could provide, if detected, evidence for SQM in the form of a strange solar system.

Our assumption of the existence of  strangelet crystal planets are based on the following possible scenarios: 
first, after the birth of strange quark stars (hot and highly turbulent environment), 
they may eject low--mass quark nuggets. It has been suggested that ejection of planetary 
clumps may happen simultaneously due to the strong turbulence of the strange 
star surface~\citep{ren2003ultra, xu2006probe, horvath2012nature}. Thus, 
the strange planetary system can be formed directly. Second, the contamination processes 
during the supernova explosion that give birth to a strange star, if the planets of the progenitor 
star can survive the violent process, then they may be contaminated by the abundant strange 
nuggets ejected from the new strange star and then be converted to strange 
planets~\citep{geng2015coalescence, wolszczan1992planetary}. If these planets are remnants 
of the progenitor star, then there is a possibility that they can be 
strange planets~\citep{caldwell1991evidence, kettner1995structure, madsen1999physics}.
Third, stellar and planetary strange--matter objects could be a remnant of a quark 
phase in the primordial universe, that may have survived until now~\citep{cottingham1994brown};
such objects could be very numerous and they can be captured by strange stars 
(or neutron stars) to form planetary systems~\citep{chandra2000dynamical}. Finally, it has been suggested in \cite{caldwell1991evidence} that a high enough cosmic ray flux of strangelets, produced by strange stars mergers, would imply that all neutron stars should be strange stars as they would have been contaminated by the influx of strangelets. 
More recently, \cite{bauswein2009mass} have performed a series of simulation in which they found that such implication (contamination of all NS by strangelets) depend on the value of the bag constant – with smaller values ($B \sim 60$ MeV/fm$^3$) associated with higher strangelet fluxes - thus making the existence of neutron stars unlikely; whereas higher values of $B$ ($ \sim 80$ MeV/fm$^3$) leads to smaller fluxes thus allowing for the co-existence of neutron and strange stars. In relation to our work - we have adopted $B \sim 60$ MeV/fm$^3$ which means that in our scenario, most likely, neutron stars would all be strange stars. There are some caveats, however. Unlike the work of \cite{bauswein2009mass}, where they have set $m_s = 100$ MeV, we have used $m_s = 150$ MeV - although we believe that this unlikely to change qualitatively the results; second and more important: such conclusions are only valid if the strange stars mergers are the only source of strangelet ejection, as pointed out in \cite{bauswein2009mass}. 


We intend to pursue further the idea of  strangelet crystal planets by adding further sophistication to the model - like a better description of curvature and surface tension effects, exploring their transport and thermal properties, and studying their seismic properties. We are currently investigating the effect of different values for the Bag constant on the properties of these objects - this would allow us to investigate the small strangelet flux scenario - as discussed in \cite{bauswein2009mass} - in which neutron stars and strange stars could potentially coexist. Nonetheless we believe that the idea set forth in this work gives rise to interesting possibility in the already rich study of strange planets and their possible observable signatures. 

%
%
\vspace*{5mm}
\section*{Acknowledgements}
J.Z. acknowledges financial support from CAPES. R.N. acknowledges financial support from CAPES and CNPq, as well as  that this work is a 
part of the project INCT-FNA Proc. No. 464898/2014-5.
Finally, J.Z. and R.N. would like to express their gratitude to the anonymous referee, whose comments have contributed considerably to improve our work.

\bibliography{biblio}

\end{document}